\documentclass[prd,onecolumn,amsfonts]{revtex4}

\usepackage{epsfig}
\usepackage{graphics}
\usepackage{amsmath}
\usepackage{epsfig}
\usepackage{graphics}
\usepackage{amsmath}
\usepackage{amssymb}
\usepackage{relsize}
\usepackage{color}

\newcommand{\p}[1]{\phantom{#1}}
\newcommand{\be}{\begin{equation}}
\newcommand{\ee}{\end{equation}}

\begin{document}

\title{Propagation of Gravitational Waves in Generalized TeVeS}
\author{Eva Sagi}
\affiliation{Racah Institute of Physics, Hebrew University of
Jerusalem, Jerusalem 91904, Israel} \email{eva.sagi@mail.huji.ac.il}

\date{\today}
\begin{abstract}
Efforts are underway to improve the design and sensitivity of
gravitational waves detectors, with the hope that the next
generation of these detectors will observe a gravitational wave
signal. Such a signal will not only provide information on dynamics
in the strong gravity regime that characterizes potential sources of
gravitational waves, but will also serve as a decisive test for
alternative theories of gravitation that are consistent with all
other current experimental observations. We study the linearized
theory of the tensor-vector-scalar theory of gravity (TeVeS) with
generalized vector action, an alternative theory of gravitation
designed to explain the apparent deficit of visible matter in
galaxies and clusters of galaxies without postulating yet undetected
dark matter. We find the polarization states and propagation speeds
for gravitational waves in vacuum, and show that in addition to the
usual transverse-traceless propagation modes, there are two more
transverse modes and two trace modes. Additionally, the propagation
speeds are different from $c$.
\end{abstract}
\maketitle
\section{Introduction}

Gravitational wave (GW) science is raising in popularity, with
several GW detectors in operation in the United States (LIGO),
Europe (VIRGO and GEO) and Japan (TAMA), and efforts under way to
improve their design and increase their
sensitivity~\cite{GW-review}. LISA, the laser interferometer space
antenna, is expected to fly in the next decade, with LISA pathfinder
marking the way~\cite{LISA}. The detection of gravitational waves
will convey new information on dynamics of systems in the
strong-field gravity limit, as well as in regions of the universe
which are opaque to electromagnetic radiation. Gravitational waves
will also serve as a test for alternative theories of gravity.
Despite the outstanding success of General Relativity (GR) in the
solar system, it fails to explain dynamics on galaxy and galaxy
cluster scales without postulating large amounts of yet-undetected
Dark Matter (DM). An alternative theory of gravitation that does not
require DM is TeVeS, the tensor-vector-scalar theory of gravity,
which was designed as a relativistic implementation of Milgrom's
MOdified Newtonian Dynamics paradigm (MOND). MOND explains the
asymptotical flatness of galaxy rotation curves without postulating
the existence of yet undetected dark matter, as well as the
sharpness of the Tully-Fisher relation which correlates luminosity
of a disk galaxy with its asymptotic rotational velocity in a
natural context. Milgrom proposed that Newtonian gravity
progressively fails as accelerations drop below a characteristic
scale $\mathfrak{a}_{0}\simeq 10^{-10}\textrm{m}/\textrm{s}^2$ which
is typical of galaxy outskirts, and assumes that for accelerations
of order $\mathfrak{a}_0$ or well below it, the Newtonian relation
$\mathbf{a}=-\mathbf{\nabla}\Phi_N$ is replaced by \be
\tilde{\mu}\left(|\mathbf{a}|/\mathfrak{a}_0\right)\mathbf{a}=-\mathbf{\nabla}\Phi_N,
\ee where the function $\tilde \mu(x)$ smoothly interpolates between
$\tilde\mu(x)=x$ at $x\ll 1$ and the Newtonian expectation  $\tilde
\mu(x)=1$ at $x\gg 1$. This relation with a suitable standard choice
of $\tilde\mu(x)$ in the intermediate range has proved successful
not only in justifying the flatness of galaxy rotation curves in
regions where acceleration scales are much below $\mathfrak{a}_0$,
but also in explaining detailed shapes of rotation curves in the
inner parts in terms of the directly seen mass, and in giving a
precise account of the observed Tully-Fisher law , $L\propto V_c^4$,
predicting the relation $L = (Ga_0\Upsilon)^{-1}V^4_c .$ This sharp
relation, while obtained naturally in the framework of MOND,
requires quite a fine tuning of dark halo parameters to be explained
by the dark matter paradigm~\cite{DM-fine-tuning}.

However, MOND alone is only a phenomenological prescription that
does not fulfill the usual conservation laws, nor does it make clear
if the departure from Newtonian physics is in the gravity or in the
inertia side of the equation $\mathbf{F}=m\mathbf{a}$. Moreover, it
is non relativistic, and as such it does not teach us how to handle
gravitational lensing or cosmology in the weak acceleration regimes.
To address these issues, Bekenstein designed the
Tensor-Vector-Scalar theory of gravity (TeVeS)~\cite{BekPRD}, a
covariant field theory of gravity which has MOND as its low
velocity, weak acceleration limit, while its nonrelativistic strong
acceleration limit is Newtonian and its relativistic limit is GR.
TeVeS sports two metrics, the ``physical'' metric on which all
matter fields propagate, and the Einstein metric which interacts
with the additional fields in the theory: a timelike dynamical
vector field, $A^{\alpha}$, and a scalar field, $\phi$. The theory
also involves a free function $\mathcal{F}$, a length scale $\ell$,
and two positive dimensionless constants $k$ and $K$. The scalar
field in TeVeS provides the additional gravitational potential for
matter, whereas the vector field provides the desired light bending
properties, in a fashion similar to the constant unit vector in
Sanders' stratified theory~\cite{SandersStratified}. The TeVeS
equations are derived from an action principle, thus ensuring
conservation of energy and momentum.

One aspect of gravitational waves in TeVeS has been investigated by
Sotani, who calculated oscillation spectra of neutron stars in the
theory generated by perturbations of the fluid~\cite{Sotani1}, as
well as by perturbations of the tensor and scalar
fields~\cite{Sotani2}, and compared them to the oscillation
frequencies predicted in GR. Recently, Sagi~\cite{sagi:044032}
calculated the PPN parameters for TeVeS and showed that there is a
link between the cosmological value of the scalar field, $\phi_c$,
and the coupling constant of the vector field, $K$, which prevents
the scalar field from evolving with cosmological time as predicted
by its equation of motion. This adds to existing evidence of
dynamical problems in TeVeS with Maxwell-like vector action, that
has been provided by Seifert~\cite{seifert} and by Contaldi et
al.~\cite{Contaldi}. Consequently, in this work we investigate
gravitational waves in vacuum for the generalized version of TeVeS
introduced by Skordis~\cite{skordis2008}, and determine the complete
set of mode speeds and polarizations for generic values of the free
parameters in the theory. This version of TeVeS has a vector action
of the most general form which is quadratic in derivatives of the
vector fields, and scalar and metric actions as in the original
formulation of TeVeS, thus preserving the correct MOND and Newtonian
limits. The vector field has four coupling constants associated with
it, instead of just one in the original version of TeVeS. The
results presented in this paper will be useful in determining the
predictions of TeVeS with regard to gravitational wave emission from
astrophysical sources.

\section{Generalized TeVeS Equations}

The metric, matter and scalar actions are given by~\cite{BekPRD}

\begin{align} \label{HEaction} &S_g=\frac{1}{16\pi G}\int
g^{\alpha\beta}R_{\alpha\beta}\, \sqrt{-g}\,d^4x,
\\
&S_m=\int
\mathcal{L}\left(\tilde{g}_{\mu\nu},f^\alpha,f^\alpha_{;\mu},\cdot\cdot\cdot\right)\,
\sqrt{-\tilde{g}}\,d^4x, \label{matteraction}\\
\label{scalaraction}&S_s=-\frac{1}{2 k^2 \ell^2 G}\int
\mathcal{F}\left(k
\ell^2\mathbf{h}^{\alpha\beta}\phi_{,\,\alpha}\phi_{,\,\beta}\right)\,\sqrt{-g}\,d^4x,
\end{align} Above $\mathbf{h}^{\alpha\beta}\equiv g^{\alpha\beta}-A^\alpha A^\beta$
with $A^\alpha\equiv g^{\alpha\beta}A_\beta$. In the scalar's action
$k$ is a dimensionless positive parameter while $\ell$ is a constant
with the dimensions of length, and ${\cal F}$ a dimensionless free
function. $G$ is the gravitational coupling constant, and is not
equal to the measured Newton's constant, which will be denoted
$G_N$. The metric which couples to the gravitational fields is
$g_{\mu\nu}$, dubbed here the Einstein metric. The matter fields,
which include all standard model particles, couple to the physical
metric, which is composed of the three gravitational fields
\be\label{grelation}
\tilde{g}_{\alpha\beta}=e^{-2\phi}g_{\alpha\beta}-2A_\alpha A_\beta
\sinh(2\phi). \ee

The vector action is taken to be of the most general form quadratic
in derivatives of the vector fields, as follows:
\begin{widetext} \be S_v=-\frac{1}{16\pi G}\int \sqrt{-g}d^4x
\left(\frac{K}{2}F_{\alpha\beta}F^{\alpha\beta}+\frac{K_+}{2}S_{\alpha\beta}S^{\alpha\beta}+K_2\left(\nabla
A\right)^2+K_4 \dot{A}_\alpha\dot{A}^\alpha-\lambda\left(A^\alpha
A_\alpha +1\right)\right)\ee \end{widetext} where
$F_{\alpha\beta}=A_{\alpha;\beta}-A_{\beta;\alpha}$,
$S_{\alpha\beta}=A_{\alpha;\beta}+A_{\beta;\alpha}$ and
$\dot{A}^\alpha=A^\beta A^\alpha_{;\beta}$. The $K_i$ are
dimensionless coupling constants. K is the coupling constant of the
original version of TeVeS.

Variation of the action with respect to $g^{\alpha\beta}$ yields the
TeVeS Einstein equations for $g_{\alpha\beta}$ \be\label{metric_eq}
G_{\alpha\beta}=8\pi G\left(
\tilde{T}_{\alpha\beta}+\left(1-e^{-4\phi}\right)A^\mu
\tilde{T}_{\mu(\alpha}A_{\beta)}+\tau_{\alpha\beta}\right)+\theta_{\alpha\beta},
\ee where $v_{(\alpha} A_{\beta)}\equiv  v_\alpha A_\beta + A_\alpha
v_\beta$, etc. The sources here are the usual matter energy-momentum
tensor $\tilde{T}_{\alpha\beta}$ (related to the variational
derivative of $S_m$ with respect to $\tilde g^{\alpha\beta}$), as
well as  the energy-momentum tensors for the scalar and vector
fields,

\begin{widetext}
\begin{align}
\label{tau} &\tau_{\alpha\beta}\equiv
\frac{\mu(y)}{kG}\left(\phi_{,\, \alpha}\phi_{,
\,\beta}-A^\mu\phi_{, \mu}A_{(\alpha}\phi_{,
\,\beta)}\right)-\frac{\mathcal{F}(y)
 g_{\alpha\beta}}{2k^2 \ell^2 G} \\
&\theta_{\alpha\beta} \equiv \,K\left(F_{\sigma\alpha}F^{\sigma}_{\p{\sigma}\beta}-\frac{1}{4} F^2 g_{\alpha\beta}\right) + {K_+} \left( S_{\alpha\sigma}S_{\beta}^{\p{\beta}\sigma} - \frac{1}{4} S^2 g_{\alpha\beta} + \nabla_\sigma \left[  A^{\sigma} S_{\alpha\beta}-S^{\sigma}_{\p{\sigma} (\alpha} A_{\beta )} \right]\right)\nonumber\\
&+K_2 \left(g_{\alpha\beta} \nabla_\sigma \left( A^\sigma \nabla \cdot A \right) - A_{(\alpha}\nabla_{\beta)} \nabla \cdot A  - \frac{g_{\alpha\beta}}{2} (\nabla\cdot A)^2 \right)\nonumber\\
& + K_4 \left(\dot{A}_\beta \dot{A}_\alpha +  \dot{A}_\sigma
A_{(\alpha} \nabla_{\beta)} A^\sigma - \nabla_\sigma
\left[\dot{A}^\sigma  A_\alpha A_\beta \right]-
\frac{g_{\alpha\beta}}{2} \dot{A}_\sigma \dot{A}^\sigma \right)-
\lambda A_\alpha A_\beta\label{new_stress_tensor},
\end{align}
\end{widetext}
where $v_{[\alpha} A_{\beta]}\equiv  v_\alpha A_\beta - A_\alpha
v_\beta$, etc.,  and \be \mu(y)\equiv\mathcal{F}'(y);\qquad  y\equiv
k\ell^2 \mathbf{h}^{\gamma\delta} \phi_{,\,\gamma}\phi_{,\,\delta}.
\ee Each choice of the function $\mathcal{F}(y)$ defines a separate
TeVeS theory. Its derivative $\mu(y)$ functions somewhat  like the
$\tilde{\mu}$ function in MOND. For $y>0$, $\mu(y)\simeq 1$
corresponds to the high acceleration, i.e., Newtonian, limit, while
the limit $0<\mu(y)\ll 1$ corresponds to the deep MOND regime.  In
the MOND regime, $\mu(y)\sim \sqrt{y/D}, $ with $D$ a dimensionless
constant. We shall only consider functions such that $\mathcal{F}>0$
and $\mu>0$ for either positive or negative arguments.

The equations of motion for the vector and scalar fields are
obtained by varying the  action with respect to $\phi$ and
$A_\alpha$, respectively.  We have \be \left[\mu(y)
\mathbf{h}^{\alpha\beta}\phi_{,\,\alpha}\right]_{;\,\beta}
=kG\left[g^{\alpha\beta}+\left(1+e^{-4\phi}\right)A^\alpha
A^\beta\right] \tilde{T}_{\alpha\beta}\,, \label{scalar_eq} \ee for
the scalar and \begin{align} &K\nabla_\alpha F^{\alpha\beta} + {K_+}
\nabla_\alpha S^{\alpha\beta} + K_2 \nabla^\beta \left( \nabla\cdot
A\right) - K_4 \dot{A}^\sigma \nabla^\beta A_\sigma + K_4
\nabla_\sigma \left(\dot{A}^\beta A^\sigma\right) + \lambda A^\beta
+ \frac{8\pi}{k}\mu  A^\alpha \phi_{,\alpha} g^{\beta\gamma}
\phi_{,\gamma} \nonumber\\& =8\pi G \left( 1-e^{-4\phi}\right)
g^{\beta\alpha}\tilde{T}_{\alpha\gamma} A^{\gamma}\label{new_vector}
\end{align} for the vector. Additionally, there is the normalization
condition on the vector field \be \label{normalization} A^\alpha
A_\alpha=g_{\alpha\beta}\,A^\alpha A^\beta=-1. \ee The $\lambda$ in
Eq.~(\ref{new_vector}), the lagrange multiplier charged with the
enforcement of the  normalization condition, can be calculated from
the vector equation.

\section{Metric, vector and scalar perturbations on a curved
background in TeVeS}

We start by considering metric, vector and scalar perturbations on a
curved background. One can derive results on the structure of the
solutions of the TeVeS equations, following the accounting system
described in~\cite{Maggiore}, Ch. 2. In order to discern the
perturbation from the background we assume that in some coordinate
system we can write the metric, vector and scalar as
\begin{align}g_{\alpha\beta}&=\bar{g}_{\alpha\beta}+h_{\alpha\beta}\label{metric_perturbation}\\
A^{\alpha}&=\bar{A}^{\alpha}+u^{\alpha}\label{vector_perturbation}\\
\phi&=\phi_B+\delta\phi,\label{scalar_perturbation}\end{align} where
$\bar{g}_{\alpha\beta},$ $\bar{A}^{\alpha}$ and $\phi_B$ have a
typical scale of variation $L_B$, on top of which small amplitude
perturbations are superimposed, characterized by a scale $\ell_g$
satisfying $\ell_g\ll L_B$ (alternatively, the distinction can be
made in frequency space, with the background characterized by a
frequency much lower that the perturbation). Additionally, we assume
that the background metric, vector and scalar are $O(1)$, whereas
the perturbations are of order $\epsilon\ll 1$. Writing the vacuum
TeVeS Einstein equations in the form \be \label{TeVeSEqRmunu}
R_{\alpha\beta}=\left(8\pi G\tau_{\mu\nu}+\theta_{\mu\nu}\right)
(\delta_{\alpha}^{\mu}\delta_{\beta}^{\nu}
-\frac{1}{2}g_{\alpha\beta}g^{\mu\nu}),\ee and expanding the Ricci tensor to $O(\epsilon^2)$, we can split the TeVeS
equations into two parts: a low frequency, long wavelength part
which describes how the background is affected by the perturbations,
and a high frequency, short wavelength part which describes the
propagation of the perturbations on the background, as follows
\be\label{lowF}\bar{R}_{\alpha\beta}=-[R^{(2)}_{\alpha\beta}]^{Low}+\left(\left(8\pi
G\tau_{\mu\nu}+\theta_{\mu\nu}\right)
(\delta_{\alpha}^{\mu}\delta_{\beta}^{\nu}
-\frac{1}{2}g_{\alpha\beta}g^{\mu\nu})\right)^{Low}\ee and
\be\label{highF}{R}^{(1)}_{\alpha\beta}=-[R^{(2)}_{\alpha\beta}]^{High}+\left(\left(8\pi
G\tau_{\mu\nu}+\theta_{\mu\nu}\right)
(\delta_{\alpha}^{\mu}\delta_{\beta}^{\nu}
-\frac{1}{2}g_{\alpha\beta}g^{\mu\nu})\right)^{High}\ee with
$\theta_{\mu\nu}$ given by Eq.~(\ref{new_stress_tensor}) and
$\tau_{\mu\nu}$ by Eq.~(\ref{tau}). In the above, $\bar{R}_{\alpha\beta}$ is constructed from $\bar{g}_{\alpha\beta}$, and contains only low frequency modes. $R^{(1)}_{\alpha\beta}$  is linear in $h_{\alpha\beta}$, and is thus high frequency, whereas $R^{(2)}_{\alpha\beta}$, which is quadratic in $h_{\alpha\beta}$, can contain both low-frequency modes generated from terms with nearly equal but opposite high wave-vectors, as well as high-frequency modes. In the absence of matter sources, the equations of motion force the amplitude of the
perturbations $\epsilon$ to be equal to $\ell_g/L_B$ (this is true only in the absence of sources; if sources are present then they determine the background curvature whereas $\epsilon\ll \ell_g/L_B$). This comes about since the Ricci tensor contains terms which are quadratic in derivatives of the metric, so that on the left hand side
of Eq.~(\ref{lowF}) we have terms quadratic in derivatives of the background metric, which are proportional to $1/L_B^2$, and on the right hand side we have terms quadratic in derivatives of the perturbation, which are proportional to $(\epsilon/\ell_g)^2$.
One can then set $L_B=1$ and use a single expansion parameter
$|h_{\alpha\beta}|,|u^\alpha|,|\delta\phi|\sim \epsilon
\sim\ell_g\ll 1.$ From Eq.~(\ref{lowF}) we see that terms quadratic
in the perturbations induce changes in the background, whereas the
propagation equations, Eq.~(\ref{highF}), are first order in the
perturbation. The leading term in the propagation equations, which
contain second derivatives of the perturbations, is then
$O(1/\epsilon)$, and first derivatives of the perturbations or of
the background metric are $O(1)$.

Regardless of the form of $\mu(y),$ the scalar equation separates
from the vector-tensor equations, since for all $y$ $0<\mu(y)\leq 1$
and $0<\mathcal{F}\leq y$. We can thus evaluate the order of
magnitude of $\tau_{\alpha\beta}$ by taking $\mu=1$ and
$\mathcal{F}= y$; this will give us a bound from above:
$\tau_{\alpha\beta} \sim \phi_{,\alpha}\phi_{,\beta}$.   Then
$\phi_{,\alpha}\phi_{,\beta} \sim O((\epsilon/\ell_g)^2)=O(1).$ The
contribution of the scalar field to the vector equation,
$\frac{8\pi}{k}\mu A^\alpha \phi_{,\alpha} g^{\beta\gamma}
\phi_{,\gamma},$ is of the same order of magnitude. Hence the scalar
contributions to the equations for the vector and the metric are
$O(1)$, and are an order $\epsilon$ smaller than the leading
contributions, containing second derivatives of perturbations.
Consequently, the vector-metric system separates from the scalar
equation on a curved background, and one can treat the two
separately.

\subsection{Vector-metric perturbations}\label{subsec_vector_metric}

We are interested in the lowest order terms in the metric and vector
equations, which are $O(1/\epsilon)$. To this order, we can
approximate $\bar{g}_{\alpha\beta}$ by the Minkowski metric
$\eta_{\alpha\beta}$, and the background scalar and vector fields by
$\bar{A}^\alpha\approx (1,0,0,0)$ and $\phi_B=\phi_c\approx const.$
This approximation is valid only in the absence of matter, where the
background curvature can be assumed to be almost absent. To avoid
carrying factors of $e^{2\phi_c}$ throughout the calculation, we
chose coordinates in which the background Einstein metric is
Minkowski; at the end of the calculation we will switch to
coordinates in which the background physical metric is Minkowski. To
first order in the perturbations, the difference between the two
coordinate systems is only one of scale. For simplicity, we elected
to work in the reference frame in which the vector field is at rest;
when including matter content, the velocity of the matter frame with
respect to the vector frame, $v$, would have to be accounted for,
and the vector field would acquire a temporal component of the order
of $v^2$ and a spatial component of the order of $v$. We will work
in units in which the speed of light is unity.

We substitute
Eqs.~(\ref{metric_perturbation},\ref{vector_perturbation},\ref{scalar_perturbation})
into the metric and vector TeVeS field equations in vacuum,
Eq.~(\ref{TeVeSEqRmunu}) and \be K\nabla_\alpha F^{\alpha\beta} +
{K_+} \nabla_\alpha S^{\alpha\beta} + K_2 \nabla^\beta \left(
\nabla\cdot A\right) - K_4 \dot{A}^\sigma \nabla^\beta A_\sigma +
K_4 \nabla_\sigma \left(\dot{A}^\beta A^\sigma\right) + \lambda
A^\beta + \frac{8\pi}{k}\mu  A^\alpha \phi_{,\alpha} g^{\beta\gamma}
\phi_{,\gamma}=0\label{vacuum_vector}.\ee  To order $O(1/\epsilon),$
indices are raised and lowered with $\eta_{\alpha\beta}$, so that
for example $u^i=u_{i}$, etc.

The temporal component of the vector equation gives $\lambda$, which
is first order in the perturbation, since the background lagrange
multiplier is zero. The spatial components are \be K_2 h_{jj,0i}+2
K_+ h_{ij,j0}+\left(K+K_+-K_4\right)(h_{00,0i}-2u^i_{,00})+2 (K+K_+)
u^i_{,jj}=0\ee

The metric equations are: \begin{align} 00:&
h_{ij,ij}-h_{ii,jj}=2(K+K_+-K_4)\left(h_{0i,0i}-\frac{1}{2}h_{00,ii}+u^{i}_{,0i}\right)\\
 0i:&
h_{0j,ji}+h_{ij,j0}-h_{0i,jj}-h_{jj,i0}=K_2h_{jj,i0}+2K_+(h_{ij,j0}+u^i_{,jj})+2(K_2+K_+)u^j_{,ji}\\
ij:& h_{k(i,j)k}-h_{0(j,i)0}-h_{,ij}-\Box
h_{ij}-\delta_{ij}\left(h_{00,00}+h_{kl,kl}-\Box
h\right)=K_{+}\left(h_{ij,00}+u^j_{,i0}+u^i_{,j0}\right)+K_2\delta_{ij}(h_{kk,00}+2u^k_{,k0})\end{align}
Round brackets denote symmetrization without a factor $1/2$. $h$ is
the trace of the metric perturbation, and $\Box$ is the flat space
d'Alembertian.  The temporal component of the vector field
perturbation is determined from the normalization condition: \be
u^0=\frac{1}{2}h_{00}\ee This system of equations is very similar to
the system obtained in {\AE}ther linearized theory, therefore we
will follow the analysis in ~\cite{AEther-waves}.

As in GR, after choosing a frame within which the metric, vector and
scalar fields have the form~(\ref{metric_perturbation},
\ref{vector_perturbation}, \ref{scalar_perturbation}), we are left
with a residual gauge symmetry. Under an infinitesimal
transformation of coordinates \be x^\mu\rightarrow
x'^\mu=x^\mu+\xi^\mu(x),\label{infi_gauge}\ee the fields transform
as \begin{align} h'_{\mu\nu}=&h_{\mu\nu}-\left(\partial_\mu
\xi_\nu+\partial_\nu
\xi_\mu\right),\\
u'^\mu=&u^\mu+\partial_0\xi^\mu,\\\delta\phi'=&\delta\phi.\end{align}
If $|\partial_\mu \xi_\nu|\sim |h_{\mu\nu}|$, then the condition
$|h_{\mu\nu}|\ll 1$ is preserved. We thus have the freedom to
perform a linearized gauge transformation on the fields. The Lorentz
gauge usually chosen in GR is of little use to us, since the
additional terms in the equations that stem from the vector field
stress-energy tensor are not simplified in this gauge. Instead, we
choose as in~\cite{AEther-waves} to impose the four conditions
$h_{0i}=0$ and $u^i_{,i}=0$. It is easy to show that they can be
obtained from a gauge transformation of the form~(\ref{infi_gauge});
starting from an arbitrary gauge that satisfies
~(\ref{metric_perturbation}, \ref{vector_perturbation},
\ref{scalar_perturbation}), one has to elect a vector $\xi^\mu$ that
satisfies \be h'_{0i}=h_{0i}-\left( \xi_{i,0}+ \xi_{0,i}\right)=0\ee
and \be u'_i=u_i+\xi_{i,0}=0\ee Adding the vector equation and the
spatial divergence of the metric equation, one gets a Poisson
equation for $\xi_0$, and $\xi_i$ can then be found by integrating
the metric equation with respect to time.

In the gauge in which $h_{0i}=0$ and $u^i_{,i}=0$, the metric-vector
system of equations takes the form:

\begin{align} & h_{ij,ij}-h_{ii,jj}=-(K+K_+-K_4)h_{00,ii}\label{time-time}\\
&h_{ij,j0}-h_{jj,i0}=K_2h_{jj,i0}+2K_+(h_{ij,j0}+u^i_{,jj})\label{time-space}\\
&h_{k(i,j)k}-h_{,ij}-\Box
h_{ij}-\delta_{ij}\left(h_{00,00}+h_{kl,kl}-\Box
h\right)=K_{+}\left(h_{ij,00}+u^j_{,i0}+u^i_{,j0}\right)+K_2\delta_{ij}h_{kk,00}\label{space-space}\\
&K_2 h_{jj,0i}+2 K_+
h_{ij,j0}+\left(K+K_+-K_4\right)(h_{00,0i}-2u^i_{,00})+2 (K+K_+)
u^i_{,jj}=0\label{vector}\end{align} we now have thirteen equations
and nine unknown functions (the divergence condition on the spatial
part of the vector field leaves only two independent vector
components), hence four of the equations above are redundant, and
will serve as checks for our calculation. We will work with the six
equations~(\ref{space-space}) and with the three vector
equations~(\ref{vector}), since they are the equations controlling
the dynamics of the system, whereas the rest of the equations are
constraint equations. To solve the system, we assume plane wave
solutions for the perturbations, in coordinates such that the wave
vector is $(k_0,0,0,k_3)$:
\begin{align}
h_{\alpha\beta}&=\epsilon_{\alpha\beta}e^{ik_\mu x^\mu}\\
u^a&=\epsilon_a e^{ik_\mu x^\mu}.\end{align} The equations become:
\begin{align} u^i:&\,
k_0k_3K_+\epsilon_{i3}+k_0^2(K+K_+-K_4)\epsilon_i-k_3^2(K+K_+)\epsilon_i=0\\
u^3:&\,
K_2\epsilon_{ii}+(2K_++K_2)\epsilon_{33}+(K+K_+-K_4)\epsilon_{00}=0
\\ E_{11}:&\,
k_3^2\epsilon_{00}+((1+K_2)k_0^2-k_3^2)\epsilon_{22}+(2K_++K_2)k_0^2\epsilon_{11}+(1+K_2)k_0^2\epsilon_{33}=0\\
E_{22}:
&\,k_3^2\epsilon_{00}+((1+K_2)k_0^2-k_3^2)\epsilon_{11}+(2K_++K_2)k_0^2\epsilon_{22}+(1+K_2)k_0^2\epsilon_{33}=0\\
E_{33}:
&\,(1+K_2)(\epsilon_{11}+\epsilon_{22})+(2K_++K_2)\epsilon_{33}=0\\
E_{12}: &\,\left((2K_+-1)k_0^2+k_3^2\right)\epsilon_{12}=0\\ E_{i3}:
&\,(2K_+-1)k_0^2\epsilon_{i3}-2K_+k_0k_3\epsilon_i=0\end{align} in
all the above $i=1,2$ and double indices imply summation.

The above equations are homogeneous in $k_\mu$, and therefore there
is no dispersion of the waves. If we define the wave speed to be
$s=k_0/k_3$ (it will be a real wave speed only if $s^2>0$), from the
requirement that the determinant of this homogeneous linear system
of equations is zero we obtain three wave speeds:
\begin{align}
s_1^2&=\frac{1}{{1-2K_+}}\nonumber\\
s_2^2&=\frac{K+K_+-2KK_+}{(1-2K_+)(K+K_+-K_4)}\nonumber\\
s_3^2&=\frac{(K_2+2K_+)(2-(K+K_+)+K_4)}{(1-2K_+)(K+K_+-K_4)(2+3K_2+2K_+)}\label{wavespeeds}\end{align}
Here we see that for $K_+=1/2$ and for $K+K_+-K_4=0$ the wave speeds
are infinite. That $K+K_+-K_4=0$ is disallowed is consistent with
Skordis` result, that $K+K_+-K_4$ is the coefficient of the time
derivative term in the vector cosmological perturbation
equation~\cite{skordis2008}, meaning that when $K+K_+-K_4=0$ there
is no growing mode in the vector field that can assist structure
formation. For values of the coupling constants $K_i$ for which
$s^2$ is positive and finite, the dispersion relation obtained is
linear, and $|s|$ represents the propagation speed of gravitational
disturbances. For these values, the theory has well defined
propagating waves solutions. If $s^2$ is negative for a mode, then
the frequency $k_0$ is imaginary, indicating the existence of
exponentially growing or decaying solutions. In such a case the
theory is unstable and hence presumably unphysical.

The fields excitations corresponding to the wave speeds
~(\ref{wavespeeds}) are:
\begin{itemize}
\item Two transverse-traceless modes corresponding to $s_1$:
$\epsilon_{22}=-\epsilon_{11}\neq 0,\epsilon_{12}\neq 0.$
\item Two transverse vector-tensor modes corresponding to $s_2$:
$\epsilon_{i}=\frac{1}{2}\sqrt{\frac{(1-2K_+)(2KK_+-(K+K_+))}{K_+(K+K_+-K_4)}}\epsilon_{i3}$,
with $i=1,2.$
\item A trace mode involving the metric trace and the vector
temporal component through the normalization condition,
corresponding to $s_3$: $\epsilon_{0}=\frac{1}{2}\epsilon_{00},$
$\epsilon_{11}=\epsilon_{22}=\frac{1}{2}(K+K_+-K_4)\epsilon_{00},$
$\epsilon_{33}=\frac{(1+K_2)(K_4-(K+K_+))}{K_2+2K_+}\epsilon_{00}.$\end{itemize}

The remaining metric equations agree with the above results. The
modes are easily classified by their different propagation speeds,
allowing us to naturally obtain the modes for the physical metric by
simply substituting our result in the expression for the physical
metric. To linear order, the physical metric is:\begin{align}
&\tilde{g}_{00}=e^{2\phi_c}(-1+h_{00}-2\delta\phi)\nonumber\\
&\tilde{g}_{0i}=-2u^i\sinh{(2\phi_c)}\nonumber\\&\tilde{g}_{ij}=e^{-2\phi_c}(\delta_{ij}(1-2\delta\phi)+h_{ij}).\end{align}
Going to Minkowski coordinates through the following coordinate
transformation \be\label{coordinate_trans} x^{\bar{0}}=e^{\phi_c}x^0
\ ,x^{\bar{j}}=e^{-\phi_c}x^j,\ee we get \begin{align}
&\tilde{g}_{00}=-1+h_{00}-2\delta\phi\nonumber\\
&\tilde{g}_{0i}=-2u^i\sinh{(2\phi_c)}\nonumber\\&\tilde{g}_{ij}=\delta_{ij}(1-2\delta\phi)+h_{ij}\label{phys_metric_minkowski}.\end{align}

We must remember to transform the wave speeds to Minkowski
coordinates as well:

\begin{align} &s_1^2=\frac{e^{-4\phi_c}}{{1-2K_+}}\nonumber\\
&s_2^2=\frac{e^{-4\phi_c}\left(K+K_+-2KK_+\right)}{(1-2K_+)(K+K_+-K_4)}\nonumber\\
&s_3^2=\frac{e^{-4\phi_c}(K_2+2K_+)(2-(K+K_+)+K_4)}{(1-2K_+)(K+K_+-K_4)(2+3K_2+2K_+)}\end{align}
We can then write the physical metric as
$\tilde{\eta}_{\mu\nu}+\tilde{h}_{\mu\nu}$, with the physical
perturbation tensor given by

\begin{align} \tilde{h}_{\mu\nu}=&\left(
\begin {array}{cccc}
0&0&0&0\\
 0&\epsilon_{11}&\epsilon_{12}&0\\
0&\epsilon_{12}&-\epsilon_{11}&0\\
0&0&0&0\\
\end {array}
\right)\cos{[\omega(t-z/s_1)]}+\left(
\begin {array}{cccc}
0&d\epsilon_{13}&d\epsilon_{23}&0\\
 d\epsilon_{13}&0&0&\epsilon_{13}\\
d\epsilon_{23}&0&0&\epsilon_{23}\\
0&\epsilon_{13}&\epsilon_{23}&0\\
\end {array}
\right)\cos{[\omega(t-z/s_2)]}\\+&\left(
\begin {array}{cccc}
\epsilon_{00}&0&0&0\\
 0&a\epsilon_{00}&0&0\\
0&0&a\epsilon_{00}&0\\
0&0&0&b\epsilon_{00}\\
\end {array}
\right)\cos{[\omega(t-z/s_3)]}\end{align}

with $
d=-\sqrt{\frac{(1-2K_+)(2KK_+-(K+K_+))}{K_+(K+K_+-K_4)}}\sinh{(2\phi_c)},$
$a=\frac{1}{2}(K+K_+-K_4)$ and
$b=\frac{(1+K_2)(K_4-(K+K_+))}{K_2+2K_+}.$

TeVeS exhibits the usual transverse-traceless propagation mode of
GR, but at a speed different from $c$. Additionally, it has two more
modes which excite both spatial and temporal directions, and one
trace mode, which is not symmetric in all spatial directions. We
will show in the next section that the scalar equation generates an
additional trace mode.

Incidentally, in the original version of TeVeS, which had all $K_i$
but $K$ zero, $s_1=s_2=e^{-2\phi_c}$, showing no dependence on the
coupling constant of the theory, whereas $s_3=0$, meaning that the
theory has no propagating scalar mode. This is consistent with the
results in ~\cite{BekPRD}.

\subsection{Scalar perturbations}\label{subsec_scalar}

Since the scalar field perturbations appear in the trace of the
physical metric, we expect the scalar field to give rise to another,
second trace mode. However, there is a subtlety in the derivation of
the lowest order contribution of the scalar equation. The scalar
equation in vacuum is \be \left[\mu(y)
\mathbf{h}^{\alpha\beta}\phi_{,\,\alpha}\right]_{;\,\beta} =0\,,
\label{scalar_eq_vacuum} .\ee To linear order in the perturbation,
$\mu(y)\approx \mu(y_B)+\mu'(y_B)(y-y_B).$ With
$\phi_B=\phi_c\approx const.$, $y_B=0$ and since for small $y$,
$\mu(y)\sim \sqrt{y}$, then $\mu'(y_B)$ diverges. Therefore, to
extract information on the propagation of scalar waves we have to
relax the assumption $\phi_B\approx const.$ and allow it to depend
on the space coordinates. This case has already been analyzed
in~\cite{BekPRD}, using the WKB approximation, as is common practice
for waves on a curved background; the scalar equation was shown to
acquire the form
\begin{align} 0&=\left(\mathbf{h}^{\alpha\beta}+2\xi H^\alpha
H^\beta\right)\delta\phi_{;\alpha\beta}\\ H^\alpha &\equiv
\frac{\mathbf{h}^{\alpha\beta}\phi_{B,\beta}}{\sqrt{\mathbf{h}^{\mu\nu}\phi_{B,\mu}\phi_{B,\nu}}}\\
\xi&\equiv \frac{d\ln{\mu(y)}}{d\ln{y}}\end{align} here terms
containing first derivatives of the scalar field perturbation were
omitted, being $O(1)$. Note that although originally this derivation
was made assuming only scalar perturbations, and ignoring vector and
metric perturbations, it is valid in the presence of metric and
vector perturbations, owing to the separation of the metric-vector
system from the scalar equation. Contributions of metric or vector
perturbations to the scalar equation are also higher order, since
there is no way of forming second-derivative terms of the metric or
the vector perturbations in the scalar equation. The wave speed was
shown to be $s_4\leq e^{-2\phi_B}$ in the deep MOND regime,
$s_4=e^{-2\phi_B}/\sqrt{2}$ in the Newtonian regime, and
$e^{-2\phi_B}/\sqrt{2}\leq s_4 \leq
\sqrt{1+2\xi}e^{-2\phi_B}/\sqrt{2}$ in the intermediate regime.
Going back to the physical metric in Minkowski coordinates,
Eq.~(\ref{phys_metric_minkowski}), we see that this generates an
additional trace mode in the physical metric, at a speed different
from that of the vector modes. In the notation of subsection
~(\ref{subsec_vector_metric}), it can be displayed as
\begin{align} \tilde{h}_{\mu\nu}= \left(
\begin {array}{cccc}
-2\delta\phi&0&0&0\\
 0&-2\delta\phi&0&0\\
0&0&-2\delta\phi&0\\
0&0&0&-2\delta\phi\\
\end {array}
\right)\cos{[\omega(t-z/s_4)]}\end{align}

\section{Conclusions}
We investigated the propagation of gravitational perturbations in vacuum for
the tensor-vector-scalar theory of gravity. We found that in the
linear approximation on a curved background, the scalar equation
separates from the vector-metric system of equations. We solved the
vector-metric system of equations to lowest order in the background
curvature, and obtained propagating wave solutions, with a linear
dispersion relation, and three distinct wave speeds depending on the
coupling constants of the theory and on the background value of the
scalar field. The corresponding physical metric perturbations can be
classified into a pair of transverse-traceless modes, another pair
of excitations of the temporal-spatial components of the physical
metric, and an asymmetric trace mode. Perturbations of the scalar
equation were ill defined to lowest order in the background
curvature; relaxing the restriction on the background scalar field,
we were able to deduce from the analysis in~\cite{BekPRD} that the
scalar field gives rise to an additional trace mode, at a different
propagation speed which depends on the background value of the
scalar field and on the free function of the theory.

TeVeS thus predicts six different modes of propagation for the
gravitational field, at four distinct speeds, all different from the
speed of light. Additionally, all speeds depend on a factor
$e^{-2\phi_B}$, which is expected to be close to unity when
$\phi_B=\phi_c$, but might induce a significant lag with respect to
the speed of light for large values of the background scalar field.
Such a lag might pose a problem for TeVeS; since the scalar and
vector fields are coupled to matter via the physical metric, one
would expect ultra high energy cosmic rays, whose velocity is close
to $c$, to emit Cherenkov-like radiation of scalar and vector
particles, if they move at a velocity higher than those scalar and
vector particles. Such an emission would cause the cosmic rays to
lose energy, and how much energy is lost would depend on the scalar
and vector particles emission rate, on the distance traveled by the
cosmic rays from their sources, and on the strength of the
matter-fields coupling. Such Cherenkov-like radiation might pose
very stringent restrictions on TeVeS' parameters, as was the case
for Einstein-{\AE}ther theory~\cite{Eliot_Moore_Stoica}.

That problem could be avoided, at least for the physical
perturbations originating in the vector-metric system, if the values
of the coupling constants of the theory were such as to make the
propagation speeds larger than $c$. The question of whether
superluminal propagation in a theory with two metrics is allowed is
still open~\cite{bruneton,Dubovsky,Eling}. If superluminal
propagation could be allowed in TeVeS without disturbing causality,
then one could also think of relinquishing the unconventional
kinetic term in the action of the scalar field, which was introduced
to prevent faster than light scalar waves.


\end{document}